\documentclass[11pt]{amsart}
\usepackage{fullpage}
\usepackage{amsthm,amsmath,amsfonts,amssymb} 

\def\dirac{\partial\mkern-9.5mu/\,}

\DeclareMathOperator{\tr}{Tr}

\renewcommand{\phi}{\varphi}

\title{Higher-derivative gauge theories \\from noncommutative geometry}
\author{Walter D. van Suijlekom}
\address{Institute for Mathematics, Astrophysics and Particle Physics,
Radboud University Nijmegen, Heyendaalseweg 135, 6525 AJ Nijmegen, The Netherlands}
\email{waltervs@math.ru.nl}
\date{11 October 2011}

\begin{document}

\begin{abstract}
In this short note we review the interpretation of the spectral action for the Yang--Mills system in noncommutative geometry as a higher-derivative gauge theory, adopting an asymptotic expansion in a cutoff parameter.  We recall our previous results on superrenormalizability of the resulting higher-derivative Yang--Mills gauge theory and confront this with \cite{ILV11} where these very results were erroneously claimed to be incorrect. We explain how their approach differs from ours, thus clarifying the apparent mismatch. 
\end{abstract}

\maketitle

The spectral action principle was introduced almost 15 years ago by Chamseddine and Connes \cite{CC96,CC97} in the context of noncommutative geometry. In the meantime, it has proven to be a very fruitful method to derive physical Lagrangians, notably the full Standard Model \cite{CCM07}. In this paper, we push these derivations further: basing ourselves on our previous \cite{Sui11b,Sui11c}, we review how one obtains superrenormalizable higher-derivative (HD) gauge theories from the spectral action, adopting an asymptotic expansion in a cutoff parameter $\Lambda$. We hope to clarify possible confusion that might follow from comparing this with \cite{ILV11}, where it was erroneously claimed that our results are in contradiction with theirs. This is mainly due to a different approach to the spectral action principle, which will be explained below.

\section{The spectral action and HD Yang--Mills theories}
The spectral action for the Yang--Mills system is the action functional given as
\begin{equation}
\label{eq:sa}
S[A;\Lambda] = \tr f(\dirac_A/\Lambda) - \tr f(\dirac/\Lambda), 
\end{equation}
with $\dirac_A$ the Dirac operator, coupled to a $SU(N)$ gauge field $A_\mu$. The function $f$ is real and even, and such that $f(\dirac_A/\Lambda)$ is traceclass, and $\Lambda$ is a real cutoff parameter. In comparison to \cite{CC96,CC97}, we have subtracted the purely gravitational term $\tr f(\dirac/\Lambda)$, focusing on the Yang--Mills part.


Already in \cite{CC96} it was shown that in four dimensions there is an asymptotic expansion as $\Lambda \to \infty$ for $S[A;\Lambda]$ such that, modulo negative powers of $\Lambda$ the spectral action is the Yang--Mills action. The remaining terms in this asymptotic expansion (as $\Lambda \to \infty$) turn out to be of interest too, appearing in \cite{CC11} and my previous work \cite{Sui11b,Sui11c}. We write the asymptotic expansion (as $\Lambda \to \infty$) as
\begin{equation}
\label{asexp}
S[A;\Lambda] \sim S_0[A] + \sum_{n>0} \Lambda^{-n} S_n[A] ,
\end{equation}
where $S_0$ is independent of $\Lambda$. 
Recall from asymptotic analysis that this means that there exists a constant $C >0$ and integer $N_0 >0$ such that for all $N > N_0$
\begin{equation}
\label{estimates}
\big| S[A;\Lambda] - \sum_{n=0}^N \Lambda^{-n} S_n[A] \big| \leq C \Lambda^{-N-1}.
\end{equation}
We interpret this in the following way:
\begin{itemize}
\item $S_0[A]$ is the physical action of interest, in this case the Yang--Mills action,
\item $\sum_{n>0} \Lambda^{-n} S_n[A]$ is a higher-derivative gauge invariant action that acts as a regulator for $S_0[A]$ as in \cite{Sla71,Sla72} (cf. also \cite[Section 4.4]{FS80}); it vanishes as $\Lambda \to \infty$.
\end{itemize}

We consider $\Lambda$ as a regularizing parameter for, in this case, Yang--Mills theory. Eventually, after a perturbative quantization and renormalization, one sends $\Lambda$ to infinity. This motivates the fact that we start with the asymptotic expansion $\sum_{n \geq 0} \Lambda^{-n} S_n[A]$, rather than $S[A;\Lambda]$ as \cite{ILV11} do. It is illustrative to compare this with lattice gauge theory where a similar approach is taken: applying a lattice regularization to the Yang--Mills action one obtains the Wilson action. One then proceeds to quantize the latter, to eventually let the lattice spacing go to zero to recover a quantization of the original theory. In our case, we turn the spectral action in a higher-derivative action that regularizes Yang--Mills theory and reduces to the Yang--Mills action as the regularizing parameter $\Lambda$ goes to infinity. 

In the example of Yang--Mills theory discussed in \cite{Sui11b,Sui11c} the inclusion of a finite number of local functionals $S_n[A]$ --- related to a choice of the function $f$ --- indeed improves the UV-behaviour of the Yang--Mills action. This is because for such HD gauge theories, the gauge field propagator (after appropriate gauge fixing) becomes essentially:
$$
\frac1{k^2 + \ldots +  \Lambda^{-N} k^{N+2}}.
$$
This clearly improves the behaviour of the usual Yang--Mills propagator which behaves as $1/k^{2}$ for large momentum $|k|$. After a careful power-counting including the interaction vertices in $\sum_{n=1}^N \Lambda^{-n} S_n[A]$ as well, one finds that the HD gauge theory is superrenormalizable. Full details of this can be found in \cite{Sui11b,Sui11c}.

In the preprint \cite{ILV11} the authors claim that these conclusions are incorrect, since they seemed to be in contradiction with their results. However, in that paper the authors consider the spectral action as given in Equation \eqref{eq:sa}, without asymptotically expanding it in large $\Lambda$. They do not interpret $\Lambda$ as a regularizing parameter and of the action as a local HD gauge theory which at lowest order (in the derivatives) is the Yang--Mills action. 
Even though both started from the same spectral action principle, there is no contradiction between our results since one is really considering different Lagrangian theories. Here, the use in \cite{ILV11} of the word weak-field is misleading: the authors claim that the above asymptotic expansion \eqref{asexp} is only valid in the weak-field approximation. As a matter of fact, the estimates for the asymptotic expansion (cf. Eq. \eqref{estimates}) do not impose any bound on $S_n[A]$, nor on the fields $A$ and their momenta. Only the (invalid) consideration of the asymptotic expansion as a convergent series expansion might require such bounds. 

\newcommand{\noopsort}[1]{}\def\cprime{$'$}

\end{document}